# Plasmonic modulator optimized by patterning of active layer and tuning permittivity

Viktoriia E. Babicheva [a,b,*] and Andrei V. Lavrinenko [a]

[a]*Department of Photonics Engineering, Technical University of Denmark, Ørsteds Plads, Bld. 343, DK-2800 Kongens Lyngby, Denmark*
[b]*Moscow Institute of Physics and Technology, Institutsky Pereulok 9, 141700 Dolgoprudny, Russia*

\* Corresponding author. Tel.: +45 45256883; fax: +45 4593 6581.
E-mail address: vbab@fotonik.dtu.dk (V.E. Babicheva)

**Abstract.** We study an ultra-compact plasmonic modulator that can be applied in photonic integrated circuits. The modulator is a metal-insulator-metal waveguide with an additional ultra-thin layer of indium tin oxide (ITO). Bias is applied to the multilayer core by means of metal plates that serve as electrodes. External field changes carrier density in the ultra-thin ITO layer, which influences the permittivity. The metal-insulator-metal system possesses a plasmon resonance, and it is strongly affected by changes in the permittivity of the active layer. To improve performance of the structure we propose several optimizations. We examine influence of the ITO permittivity on the modulator's performance and point out appropriate values. We analyze eigenmodes of the waveguide structure and specify the range for its efficient operation. We show that substituting the continuous active layer by a one-dimension periodic stripes increases transmittance through the device and keeps the modulator's performance at the same level. The dependence on the pattern size and filling factor of the active material is analyzed and optimum parameters are found. Patterned ITO layers allow us to design a Bragg grating inside the waveguide. The grating can be turned on and off, thus modulating reflection from the structure. The considered structure with electrical control possesses a high performance and can efficiently work as a plasmonic component in nanophotonic architectures.

## 1. Introduction

Recently plasmonics has grown in a highly developed and advanced field with a lot of interesting effects and useful applications [1-3]. It allows combining nanoscales of devices and sensors with very fast response, so extending their functionality over the whole optical frequencies. One of the central problems in plasmonics is localization and transfer of optical energy on a nanoscale close to an interface between media with positive and negative real parts of dielectric functions. Such waves called surface plasmon polaritons (SPPs) are strongly confined to the interface. Advanced techniques to manipulate SPPs as information carriers for integrated circuits are been developing [4-6].

Active plasmonics and plasmonics switching devices are considered as one of the most challenging directions in nanophotonics [7,8]. Various types of compact modulators that utilize SPPs have been proposed [1,9-14]. Both phase and absorption modulations have been exploited to achieve high speed and low footprint of a modulator along with the CMOS-compatible technology. The structural phase transition in vanadium dioxide $VO_2$, which exhibits strong contrast between the optical properties of its insulating and metallic phases, is analyzed for different plasmonic modes. Depending on a mode, either an index modulator (with $\Delta n > 20\%$) or absorption modulator can be utilized [15]. An absorption modulator based on a metal-semiconductor-metal structure with the gain core is proposed and analyzed in [16]. InP-based semiconductor materials allow achieving high gain and even a complete loss compensation in the structure. Electrical current through the active core controls the gain level and thus transmittance of the system.

However, most of plasmonic modulators are based on control of gap plasmonic modes by the charge accumulation layer. Two different layouts are possible. In the first design, a metal-oxide-semiconductor (MOS) stack is deposited on top of a silicon waveguide [17-22]. Such waveguide-integrated plasmonic modulators are reasonably simply to fabricate, their size is comparable with the silicon waveguide size, and propagation losses are originated only from one metal interface. An ultra-compact electro-optic modulator based on Silicon-on-Insulator waveguide with an ITO-$SiO_2$-Au stack is reported in [21]. High extinction ratio 1 dB/μm, low insertion loss -1 dB and broadband operation due to the non-resonant MOS mode are shown.

The second concept includes layers of oxide and semiconductor which are embedded in metal from both sides. Metal-insulator-metal structures keep additional advantages and possess strong ability to confine light [23-25]. In such type of waveguides light is localized in a gap with typical sizes ~100 nm and less, that facilitates manipulation of light on the subwavelength scale. The SPP propagation length in the metal-insulator-metal waveguide can be up to ten micrometers.

Horizontally arranged metal-insulator-semiconductor-insulator-metal slot waveguides exhibit high performance [26-28]. The main principle is based on inducing a highly accumulated electron layer at the SiO$_2$/Si interface. An electro-absorption CMOS-compatible modulator was characterized: 3-dB operation on 3 μm length at ~6.5 V bias and broadband modulation is achieved [26]. Such configurations can be readily integrated in standard Si circuits. However, high mode localization requires a very high aspect ratio of the waveguide core.

Ultra-compact efficient plasmonic modulators based on strong light localization in vertically arranged metal-semiconductor-insulator-metal waveguides have been studied recently [22,29-33]. Two metal surfaces serve also as electrodes, thus simplifying design. The phase modulator PlasMOStor consists of semiconductor core with Si and SiO$_2$ layers, sandwiched between two silver plates [29]. It supports both photonic and plasmonic modes, which interfere while propagating. The modulation is based on cutting off the photonic mode and thereby changing the integral transmittance.

Recently the concept of utilizing transparent conducting oxides, namely indium tin oxide (ITO), as an active material with varied carrier density has been studied [20-22, 30-34]. A unity-order refractive index change is achieved in a metal-oxide-semiconductor-metal multilayered stack [34]. A further progress is reported in [22], where an absorption modulator SPPAM based on a two-layer core of silicon nitride Si$_3$N$_4$ (or oxide SiO$_2$) and an ultrathin ITO layer is described. The carrier density in the transparent conducting oxide layer changes under modulating voltage. In turn it changes the effective plasma frequency of the layer. The authors employ analytical solutions of the SPP dispersion equation in a four-layer structure; the Thomas-Fermi screening theory to derive the carrier density distribution as well as numerical simulations with a finite element method. The structure supports SPP resonance at telecommunication wavelength 1.55 μm owing to the ITO layer which has small absolute values of permittivity in near-infrared region. The resonance is broad because of high losses in ITO. It decreases device's performance and increases bandwidth of operation at the same time. A similar structure based on a silicon-waveguide-integrated multilayer stack was fabricated and characterized. The logarithmic extinction ratio regarding to power up to 0.02 dB was achieved. Meanwhile a theoretical analysis of the Ag-ITO-Si$_3$N$_4$-Ag structure predicts the 1dB extinction ratio on 0.5 μm length. However, power transmitted through the 0.5-μm-long device is rather low; in particular losses are 24 dB/μm for the waveguide with Si$_3$N$_4$-core and 9 dB/μm with SiO$_2$-core. There is always a trade-off between modulation depth and transmittance through the modulating system. Thus, the design requires certain optimizations.

To improve transmittance we apply a periodic structuring of the ITO layer. Also, we analyze effects of varying the ITO permittivity as a result of different annealing conditions. The improved performance of the modulator is achieved regarding the reference system from [22]. The structure of the article is the following. In Section 2 we consider the surface plasmon polariton absorption modulator with one and two ITO layers. Section 3 shows the influence of the ITO permittivity under varying it on several units. In Section 4 we analyze eigenmodes of the modulator structure followed by study of the periodic patterning in the system (Section 5). Another concept based on a Bragg reflector grating is proposed for modulation in Section 6. And finally we sum up results and discuss them in Section 7.

## 2. Concept of absorption modulator

First we start with the definition of a reference system. The modulator consists of 8-nm-thick ITO layer and a 70-nm thick silicon nitride (Si$_3$N$_4$) core with $\varepsilon_{core}$ = 4 (Fig. 1a). These layers are sandwiched between two silver plates acting as electrodes. The thicknesses are considered as optimal based on the trade-off between the propagation length and performance [22]. Data for the ITO parameters are taken from [35] and [36]. We assume that the ITO permittivity is approximated by the Drude formula (see [35]):

$$\varepsilon = \varepsilon_\infty - \frac{\omega_{pl}^2}{\omega^2 + i\gamma\omega}, \qquad (1)$$

where $\varepsilon_\infty$ = 3.9, plasma frequency $\omega_{pl}$ = 2.9·10$^{15}$ s$^{-1}$ and collision frequency $\gamma$ = 1.8·10$^{14}$ s$^{-1}$. The carrier density in the ITO layer changes under applying modulating voltage between the electrodes. We follow the approach of Ref.22 to calculate these changes. It employs the Thomas-Fermi screening theory and averaging of the carrier density over the whole ITO layer. Thus, 5% increasing of the average carrier density alters the plasma frequency $\omega_{pl}$ from 2.9·10$^{15}$ to 2.9716·10$^{15}$ s$^{-1}$, that shifts the ITO permittivity from -1.67 + i0.825 to -1.95 + i0.867 at telecommunication wavelength 1.55 μm. Data [37] for a silver permittivity are approximated by the Drude formula (1), where $\varepsilon_{\infty(Ag)}$ = 1, plasma frequency $\omega_{pl(Ag)}$ = 1.38·10$^{16}$ s$^{-1}$ and collision frequency $\gamma_{(Ag)} = 3.22 \cdot 10^{13}$ s$^{-1}$. It gives $\varepsilon_{Ag} = -128.7 + 3.44i$ at $\lambda = 1.55$ μm.

A figure of merit (FoM) which characterizes a performance of the absorption modulator is defined directly through the absorption coefficients [38]:

$$\text{FoM} = \frac{|\alpha_{\text{off}} - \alpha_{\text{on}}|}{\alpha_{\text{state}}}, \quad (2)$$

where $\alpha_{\text{off}}$ and $\alpha_{\text{on}}$ are the absorption coefficients (regarding the field amplitude) in the voltage-off and voltage-on states, respectively. Coefficient $\alpha_{\text{state}}$ is a residual absorption, i.e. either $\alpha_{\text{off}}$ or $\alpha_{\text{on}}$ depending on which state is transmitting. Our FoM differs only by a constant coefficient from the one introduced in [22], which is formulated regarding the propagation lengths. Such a definition of FoM gives length-independent characteristic of the structure. A proper length of the device can be chosen according to signal level requirements and fabrication restrictions for particular geometry of a modulation problem.

We reduce dimension of the problem and consider the modulator as a two-dimensional four-layer metal-insulator sandwich, as shown in Fig. 1b. We solve numerically the SPP dispersion equation for the two-dimensional structure the same way as in [22] and [15]. The reference system supports two SPP modes in the transverse magnetic (TM) polarization.

The first mode with the lowest absorption (studied in [22]) has $\alpha_{\text{off}} = 1.08$ μm$^{-1}$, $\alpha_{\text{on}} = 0.83$ μm$^{-1}$ and corresponding FoM$_1$ = 0.30 at $\lambda = 1.55$ μm. The second mode has a very high absorption coefficient, i.e. $\alpha_{\text{off}} = 5.47$ μm$^{-1}$ and $\alpha_{\text{on}} = 5.87$ μm$^{-1}$, that gives a fairly low FoM$_2$ = 0.07. In Section 4 we analyze the second mode in more detail. However, because of such high absorption of the second mode we study performance of the device only for the first one.

We calculate also a transmission coefficient of the reference modulator. It exhibits significant attenuation of propagating waves due to the ITO layer. The transmitted signal amplitude in the voltage-off state is only 12% of the incoming wave for modulator's length $L = 2$ μm and approximately 1% for $L = 4$ μm. These numbers specify a problem which the reference design experiences, namely a rather low level of transmitted signals. Our optimization aims to increase this level.

For the purposes of further designing steps we also analyze the symmetric system: it consists of the Si$_3$N$_4$ core embedded between two 8-nm-thick ITO layers and sandwiched between silver plates (Fig. 1c). In this case the bottom layer experiences a carrier depletion and decreasing of the plasma frequency ([on–]-state), that is opposite to the top layer ([on+]-state). While reference voltage results in the 5% increasing of the average carrier density in the top ITO layer, the plasma frequency $\omega_{pl}$ for the bottom ITO layer changes from $2.9 \cdot 10^{15}$ to $2.8267 \cdot 10^{15}$ s$^{-1}$ due to the 5% decreasing of the average carrier density under the same voltage. Correspondingly, the ITO permittivity is shifted from $-1.67 + i0.825$ to $-1.95 + i0.867$ (top layer) and to $-1.39 + i0.784$ (bottom layer) at $\lambda = 1.55$ μm. In total, the whole five-layer system possesses a very low FoM$_{5\text{-layers}}$ = 0.05, i.e. the controllability of the system is lost. The effective permittivity of the three layers sandwiched between the silver plates under applied voltage is turned out to be approximately the same as without the voltage. Meanwhile, the additional ITO layer doubles maximal absorption in plasmonic resonance of the system, and the effect is even more pronounced on the resonance slope. Thus, for $\lambda = 1.55$ μm the five-layer system has approximately four times higher absorption coefficient than the reference asymmetric case, in particular $\alpha_{\text{off}} = 4.05$ and $\alpha_{\text{on}} = 4.25$. However, in Section 6 we show some advantages of using two ITO layers with patterning.

Yet another merit in the performance of the four-layer modulator can be achieved by applying bias in both directions. In this case the change of the plasma frequency for the voltage [on+]-state and [on–]-state is between $2.9716 \cdot 10^{15}$ and $2.8267 \cdot 10^{15}$ s$^{-1}$, that gives varying of absorption coefficient between $\alpha_{\text{on–}} = 1.47$ μm$^{-1}$ and $\alpha_{\text{on+}} = 0.83$ μm$^{-1}$, and corresponding FoM$_{\text{flip}}$ = 0.78. The efficiency is better than in case of applying the double voltage in one direction: the average carrier density changes on 10%, the ITO plasma frequency becomes $3.0415 \cdot 10^{15}$ s$^{-1}$, giving $\varepsilon_{\text{ITO}} = -2.23 + i0.908$, $\alpha_{\text{on}} = 0.657$ and finally FoM$_{10\%}$ = 0.64. The flipping of voltage direction can be also preferable for systems with a low break-through voltage threshold.

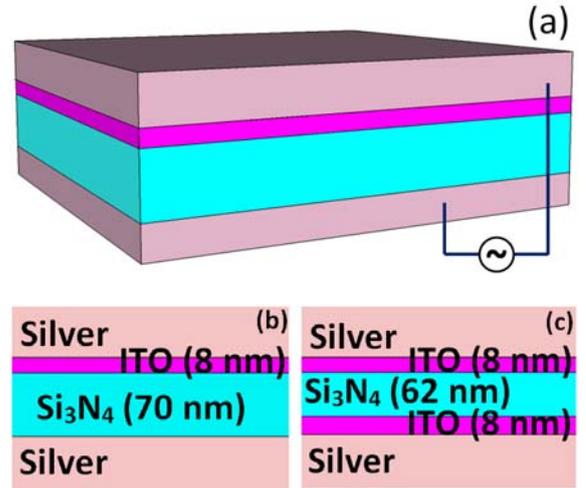

Fig. 1. a) The reference modulator structure: 8-nm ITO layer and 70-nm Si$_3$N$_4$ layer are embedded between the silver plates [22]. b) Schematic two-dimensional view of the equivalent four-layer system. c) The symmetric five-layer system with two 8-nm-thick ITO layers and 62-nm-thick Si$_3$N$_4$ core.

## 3. Effect of ITO permittivity changes

To improve modulator's performance we study variations of the ITO permittivity. In practice it is possible to vary the optical properties of transparent conducting oxides by different anneal conditions, i.e. temperature and environment, during the fabrication process. The range of the permittivity variations can be several units or even more [39]. We assign the variations in the ITO permittivity to parameter $\varepsilon_\infty$ in Drude formula (1). Thus, in our study, we sweep the real part of the permittivity keeping the change of the plasma frequency $\omega_{pl}$ (from $2.9 \cdot 10^{15}$ s$^{-1}$ to $2.9716 \cdot 10^{15}$ s$^{-1}$). We assume the collision frequency $\gamma$ being also fixed and therefore the imaginary part of the permittivity is kept as before, namely 0.825 and 0.867 for the off-state and on-state, respectively. So the difference between the off-state and on-state permittivities ($\varepsilon_{off} - \varepsilon_{on} = 0.28 - i0.042$) is fixed too.

The results of calculations for the absorption coefficients and FoM (2) are shown in Fig. 2. The FoM reaches the first maximum on the steepest slop of the absorption coefficient curve. It then nullifies in the point of equal absorption coefficients in both states and then increases again up to the biggest FoM$_{max}$ = 0.37 for Re($\varepsilon_{off}$) ≈ 0.83 that corresponds to $\varepsilon_\infty$ = 6.4. In principle such values of Re($\varepsilon_{off}$) are available with the state-of-the-art fabrication technology [39]. Consequently, our further study is carried out with this optimized value of $\varepsilon_\infty$. We note that Re($\varepsilon_{off}$) ≈ - 1.27 ($\varepsilon_\infty$ = 4.3) gives high FoM = 0.36 as well. Fig. 2 indicates that the absorption coefficient for positive Re($\varepsilon_{off}$) is higher in the on-state than in the off-state. It means that by applying voltage we decrease transmittance of the system.

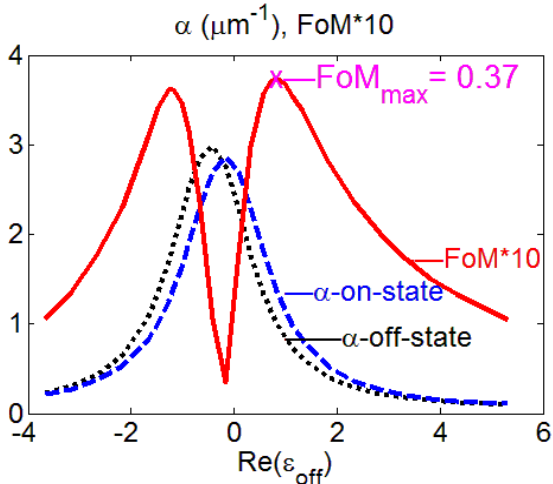

Fig. 2. The absorption coefficients α and FoM of the four-layer system and various permittivities of the ITO layer. Maximal values FoM$_{max}$ = 0.37 and FoM = 0.36 are achieved for Re($\varepsilon_{off}$) ≈ 0.83 and Re($\varepsilon_{off}$) ≈ -1.27 respectively.

Further, we analyze a five-layer system (Fig. 1c) in the situation with the reversed sign of the changes in losses. Re($\varepsilon_{off}$) of different signs can be obtained under different anneal conditions for the bottom and top ITO layers, as they are deposited at different stages of fabrication. The same way as for the four-layer system we vary Re($\varepsilon_{off}$) of the bottom layer in broad range and keep Re($\varepsilon_{off}$) ≈ 0.83 for the top layer. Calculations show that the additional layer does not provide any advantages and none of Re($\varepsilon_{off}$) can result in FoM exceeding FoM$_{max}$ = 0.37 for the optimal four-layer structure.

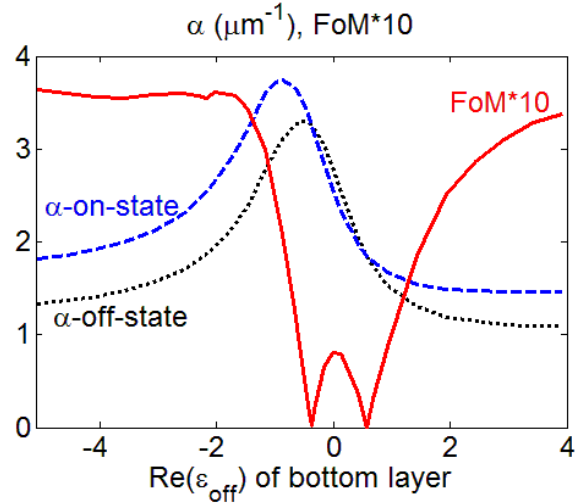

Fig. 3. The absorption coefficients α and FoM of the five-layer system with various permittivities of the bottom ITO layer. The permittivity of the top layer is fixed Re($\varepsilon_{off}$) ≈ 0.83. FoM values are not exceeding FoM$_{max}$ = 0.37 for the four-layer system.

## 4. Eigenmodes of four-layer system

In this section we analyze eigenmodes of the two-dimensional four-layer structure with the optimized ITO permittivity $\varepsilon_\infty$ = 6.4 (Fig. 1b). We solve numerically the SPP dispersion equation in the frequency range between 130 THz and 220 THz. For frequencies below 220 THz the system supports only two modes: the propagation constants and absorption coefficients are shown in Fig. 4 and Fig. 5 respectively.

We consider three states: the off-state with the ITO plasma frequency $2.9 \cdot 10^{15}$ s$^{-1}$ (without any voltage); [on+]-state under applied voltage with the ITO plasma frequency $2.9716 \cdot 10^{15}$ s$^{-1}$; and [on–]-state under the reverse bias with the plasma frequency $2.8267 \cdot 10^{15}$ s$^{-1}$. We also plot the propagation constant for the three-layer system with the 78-nm-thick Si$_3$N$_4$ core (without ITO layer).

The propagation constants of both modes exhibit monotonous increase with frequency apart from the region 160-200 THz, where $\beta$ of the first mode decreases being strongly dependent on changes in ITO permittivity. Consequently, the absorption coefficient of the first mode has maximum in that frequency region (SPP resonance). The absorption coefficient of the second mode monotonically decreases.

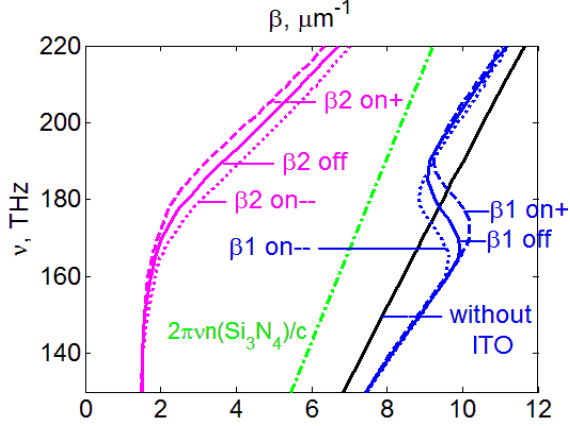

Fig. 4. Propagation constants $\beta$ for two modes "1" and "2" without voltage (off-state), under applied direct voltage ([on+]-state) and under applied reverse voltage ([on–]-state) for different frequencies. Green dot-dash line indicates the light line in $Si_3N_4$ with refractive index $n = 2$.

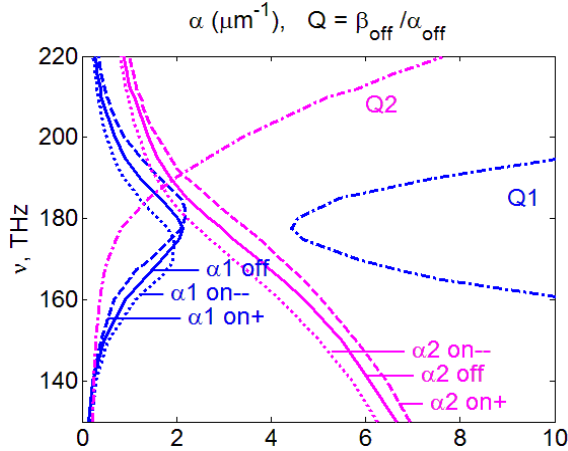

Fig. 5. Absorption coefficients $\alpha$ and quality factor $Q = \beta_{off}/\alpha_{off}$ for two modes "1" and "2" without voltage (off-state), under applied direct ([on+]-state) and reverse ([on–]-state) voltage for different frequencies.

Plots in Fig. 4 and Fig. 5 indicate that for the first mode the modulator working region should be between 160 THz and 220 THz, where a change of the ITO plasma frequency strongly affects the propagation constant and absorption coefficient of the system. Therefore, both mechanisms of modulation: the absorption change and the propagation constant variation can be applied. The latter is illustrated for a grating system further. Despite that the propagation constant has a large response on applied voltage around 180 THz, the effect causes an increase of absorption. Designing device, this circumstance should be taken into account.

Fig. 6 shows field distributions of two modes in the waveguide cross-section at frequency 160 THz. Both modes are plasmonic and quasi-symmetric (the ITO layer removes symmetry). They can be undesirably excited at the same time. However, in contrast to [21] or [29], the second mode damps much faster and consequently neither pronounced interference nor beatings can appear. We characterize the quality of a mode by factor $Q = \beta_{off}/\alpha_{off}$. Because of the high absorption and small propagation constant of the second mode, the Q-factor for the second mode is much lower than for the first one, e.g. $Q_1 \approx 10$ and $Q_2 \approx 2.5$ for 193.4 THz (see Fig. 5). Furthermore, the second mode has much lower FoM than the first one. Thus, the second mode does not support efficient performance.

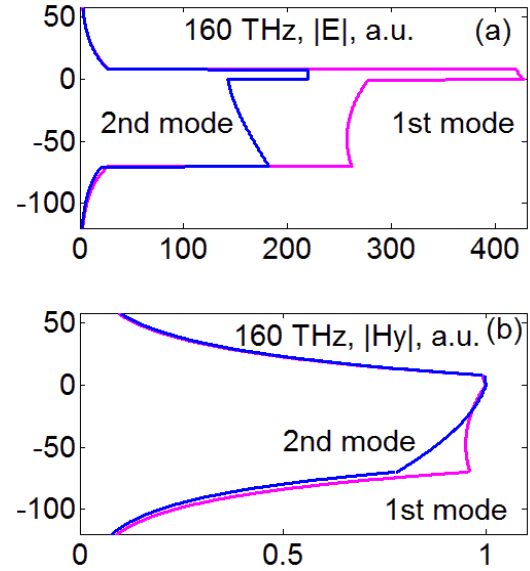

Fig. 6. Field distributions (absolute values of electric field $E$ (a) and transverse magnetic component $Hy$ (b)) of two plasmonic modes in the vertical cross-section. Calculations are performed for the off-state system. Fields are normalized such that $|Hy| = 1$ at the interface between ITO and $Si_3N_4$.

In the considered frequency range the second mode has smaller propagation constant than that of light in the dielectric core (the mode is above the light line). It allows the exponential increase of the electric field from the interface between the ITO film and dielectric core (see Fig. 6). Thus, it is a quasi-leaky mode [40], and it possesses high modal losses. Similar to [41] the system has an antenna mode region ($Q_2 > 1$) and a reactive mode region ($Q_2 < 1$).

As one can see in Fig. 5 the transition occurs at 180 THz. At the transition frequency the system changes its response on applied voltage: in the reactive region the absorption coefficient is mostly affected and the propagation constant is not, while in the antenna region it is vice versa. Calculations show that the second mode crosses the light line at 300 THz (assuming the same optical parameters of silver and ITO) and becomes a bound mode. The small propagation constant of the second mode can provide efficient coupling from free space or another waveguide.

### 5. Periodic patterning of ITO layer

The effective modulation performance means the deep modulation range and improved transmittance, which we aim implementing optimizations in the reference system. To increase transmittance we replace the continuous ITO film by periodic stripes (Fig. 7a). The patterning is characterized by a filling factor defined as the ratio of the stripe width $w$ to the period $P$ of the structure, $f = w/P$. $f = 1$ corresponds to the continuous ITO film.

The system with patterned layer requires numerical solution of Maxwell's equations. We perform simulations with the commercial software package CST Microwave Studio in the frequency domain [42]. The silver plates are 120-nm-thick, which is enough to keep the domain-termination error at a negligible level.

We analyze several systems with different period $P$ and total length $L$. Results of simulations for transmission coefficient in the voltage-off state and FoM versus the filling factor are shown in Fig. 7b ($\lambda$ = 1.55 μm). As we foresaw the transmission increases dramatically with the fall of the filling factor. The finer is sampling (less period $P$) the higher is the transmission coefficient for the same total length and filling factor. For all three periods ($P$ = 250 nm; 500 nm and 4 μm) the FoM is almost constant in the broad range of filling factors. However, for $f \leq 0.2$ FoM falls down abruptly and the fall is steeper for larger periods.

In the first approximation transmission through one period can be considered as transmission through two successive layers: a part of the waveguide with the ITO stripe (length $w$ and absorption coefficient $\alpha_{state}$), and a part of the waveguide without the ITO stripe (length $(P-w)$ and absorption coefficient $\alpha_0$). Thus, total transmission coefficient of the patterned waveguide with length $L$ is:

$$Tr_{state} = \exp(-\alpha_{state} f L - \alpha_0 (1-f) L), \quad (3)$$

where "state" is either "on" or "off", and $\alpha_0 = 0.027$ μm$^{-1}$ for 78-nm-thick Si$_3$N$_4$ core.

In this case FoM can be found as:

$$\text{FoM}(f) = \frac{|\log(Tr_{on}) - \log(Tr_{off})|}{\log(Tr_{off})} = \frac{\text{FoM}_{max}}{1 + (1-f)\alpha_0/f\alpha_{off}} \quad (4)$$

Fig. 7b shows good agreement of approximation (3) with numerical simulations. It completely coincides with results for $P$ = 4 μm and $L$ = 4 μm. Formula (4) explains the constant value of the FoM in the broad range of $f$. For $f$ = 0.2 the ratio $(1-f)\alpha_0/f\alpha_{off} \approx 0.1$, and for $f \leq 0.2$ the additional term in denominator (4) has an essential contribution. It means that losses in the part of waveguide without ITO influence the total transmission coefficient, and meanwhile the fraction of ITO is insufficient for the effective change in the propagation.

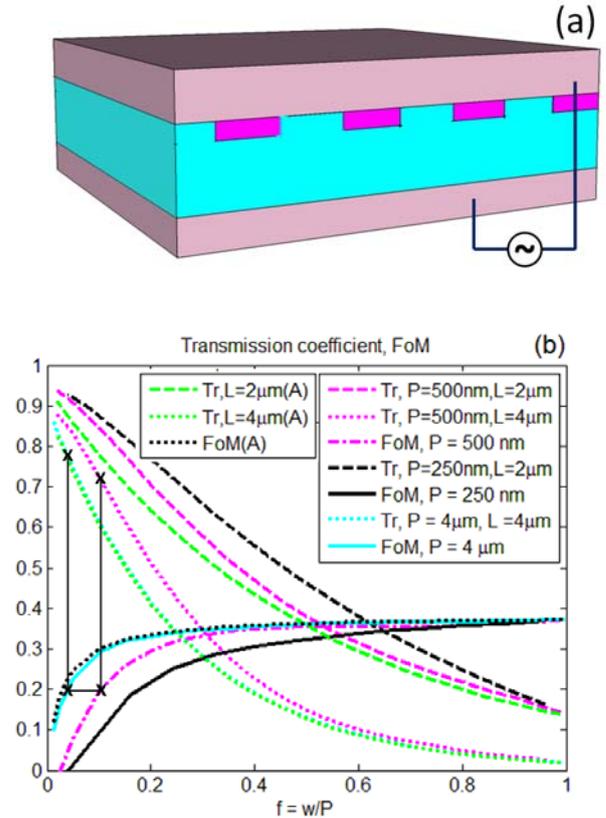

Fig. 7. a) The improved modulator design with periodic ITO stripes. b) Transmission coefficient (Tr) and FoM of the periodically patterned systems versus filling factor. Samples have different lengths ($L$ = 2 and 4 μm) and periods ($P$ = 250 nm, 500 nm and 4 μm). The structure with larger period outperforms among two patterned structures with equal FoM (marked by "x"). Transmission coefficients (A) correspond to approximation (3). FoM(A) is calculated with formula (4).

The transmission coefficient has a slight dependence on the period (for samples of equal total length *L*), and the smallest periods give highest transmission coefficient. Nevertheless, for the certain FoM the systems with highest period have highest transmission (compare two systems marked with crosses in Fig. 7b). So, partial removing of the lossy ITO layer with as big as possible period can be recommended aiming the higher transmittance with the same performance. For the following analysis we take filling factor $f$ = 0.2 as optimal patterning that gives high transmittance.

To study performance of the patterned system we consider one period of the waveguide, which contains one ITO stripe and has length *P*. Throughout numerical simulations two outbound metal corners of the stripes are rounded with the curvature of 4 nm (apart from the 5-nm stripes with the radius of curvature 2.5 nm). Such curvature value is chosen to suppress any spurious field enhancement at the sharp corners and in the same time it is not too large to cause serious deviations in the structure response. To compare transmission efficiency in the case of different periods (the filling factor is fixed) the transmission coefficients are renormalized to the chosen structure length 4 μm using the scattering matrix formalism (see e.g. [43]).

We plot amplitude transmission spectra (Fig. 8) for various sizes of the stripes from 5 nm to 800 nm as well as for the continuous homogeneous film with the dielectric function

$$\varepsilon_{\text{eff}} = f\varepsilon_{\text{ITO}} + (1-f)\varepsilon_{\text{core}}, \qquad (5)$$

where $f$ = 0.2. This dielectric function effectively approximates the permittivity of the patterned ITO layer [44].

All spectra exhibit a pronounced minimum in the transmission with clear dependence of its position on the stripe width. Spectrally the minimum dramatically shifts for stripes with sizes below 100 nm and reaches the outward-red minimum for the continuous film with permittivity $\varepsilon_{\text{eff}}$ (5).

We explain such spectral shift by excitation of a hybridized plasmonic mode on single stripe [45]. The stripe of width $w \leq 100$ nm and thickness $h$ = 8 nm with rounded corners can be considered as a nanowire with the quasi-elliptical cross section and aspect ratio $w/(2h)$. The silver plate connected to the ITO stripe serves as a mirror and effectively doubles the stripe thickness. Such quasi-elliptical nanowire in the electric field polarized along the minor axis supports dipole and quadrupole modes [45]. Their excitation causes intensive decrease in the transmission coefficient. So the overlap of their responses produces the transmission spectra we observe in Fig. 8.

To reveal the exact positions of both resonances we perform simulations for a hypothetic material identical to ITO but with diminished losses. The spectral profiles became sharper and deeper for 4-times-reduced *γ* and splitting of the peak is clear in Fig. 9a. Further decreasing of losses makes resonances even narrower. Simulations also show that for systems with fixed width $w \geq 30$ nm the positions of both dips are independent on filling factor in the range $f \leq 0.5$.

It should be mentioned that applying the corners smoothening is especially important in the simulations with reduced losses. While transmission spectra for system with realistic ITO parameters are only slightly affected by the rounded concerns of stripes, the results are strongly affected in the case of diminished losses. A system with sharp corners and small losses has larger numbers of dips in the transmission spectra. Moreover, the approximation by an elliptical nanowire is not applicable in this case.

Simulations show that one dip is almost unaffected by the stripe width variation. It is fixed near 130 THz (see Fig. 9a). It corresponds to the quadrupole mode. Another dip, which belongs to the dipole mode, red-shifts from approximately 175 THz to 150 THz with changing stripe width from 800 nm to 15 nm. By further decreasing of the stripe width the dip shift becomes more pronounced and goes even below 100 THz (see Fig. 8). Frequencies of the transmission minima are well correlated with the nanorods resonances obtained with the help of analytics from Ref. 45. Details will be presented elsewhere.

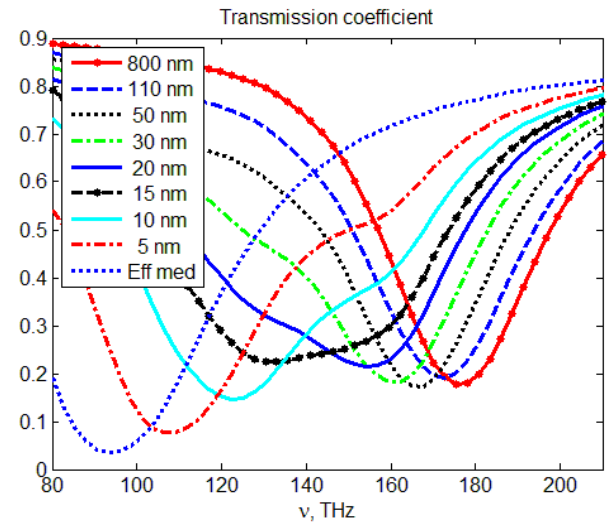

Fig. 8. The transmission spectrum has absorption maximum which shifts towards lower frequencies. Stripes have width *w* in the range from 5 nm to 800 nm. The continuous film is calculated in the effective medium approximation (5). Simulations are performed for the off-state.

Fig. 9a presents changes in the transmission spectrum for turned on bias as well. Changes are the most pronounced in the region of the highest slope of the spectra. The shift of the absorption maximum affects position of the maximal FoM. Decrease of stripe sizes shifts the maximal FoM top toward lower frequencies (Fig. 9b). Having in mind telecommunications applications (193.4 THz), the absorption maximum of the modulator has to be around 180 THz. Therefore, it is important to work with stripes above 100 nm.

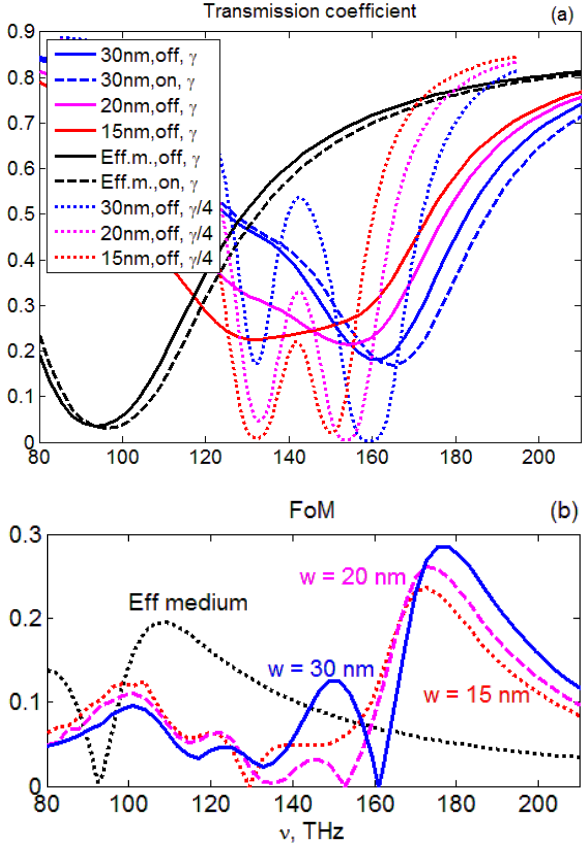

Fig. 9. (a) The amplitude transmission spectra for stripes of various widths $w$ in the off- and on-states. Decreasing of the collision frequency $\gamma$ in 4 times clearly shows the absorption maxima. (b) FoM for different stripe sizes.

## 6. Bragg grating

It is discussed in Section 4 there is a frequency range where the pronounced change in the mode propagation constants is accompanied by the absorption coefficient variations. So the absorption modulation performance can be further improved by the phase effect.

It was shown that a plasmonic Bragg reflector can be formed by the metal-insulator-metal waveguide with periodic changes of the insulator material (index-modulated reflectors) or/and insulator thickness (thickness-modulated reflectors) [46-49]. Waves propagation in the structures with various shapes (step-profile, s-shaped, sawtooth-profile, triangular-shaped and so on) is described in terms of effective indices of the guided modes. We study the system shown in the inset of Fig. 10. The bottom ITO stripes are placed periodically in anti-phase to the top stripes. We assume that there are no any gaps or overlapping between the stripes after their projection on a horizontal plane. The off-state system is supposed to be a uniform waveguide with the complex propagation constant $\beta_{\text{off}} + i\alpha_{\text{off}}$, which we can find in the first approximation from the band diagrams (Fig. 4 and 5). Therefore, we neglect all multiple reflections inside off-state structure. Under applied voltage the part of the waveguide with the top (bottom) ITO stripes supports the mode with propagation constant $\beta_{\text{on+}}$ ($\beta_{\text{on-}}$). The concept of modulation is similar to that proposed in Ref. 50: propagating waves experience Bragg reflection because of the periodic variation of effective mode indices along the waveguide under applied voltage.

The grating with the top stripe width $w_{\text{top}}$ and the bottom stripe width $w_{\text{bot}}$ obeying condition

$$w_{\text{top}}\beta_{\text{on+}} = w_{\text{bot}}\beta_{\text{on-}} = \pi/2 \qquad (6)$$

satisfies the Bragg condition and causes enhanced reflection from the system.

For a quantitative analysis of the Bragg grating performance we chose two particular frequencies: $\nu_1 = 177.5$ THz and $\nu_2 = 210$ THz. At frequency 177.5 THz $\beta_{\text{off}} = 9.4$ μm$^{-1}$, $\beta_{\text{on+}} = 10.0$ μm$^{-1}$, $\beta_{\text{on-}} = 8.9$ μm$^{-1}$, and according to (6) $w_{\text{top}} = 157$ nm and $w_{\text{bot}} = 177$ nm (the first set). At frequency 210 THz $\beta_{\text{off}} = 9.4$ μm$^{-1}$, $\beta_{\text{on+}} = 10.3$ μm$^{-1}$, $\beta_{\text{on-}} = 10.6$ μm$^{-1}$, and correspondingly $w_{\text{top}} = 152$ nm and $w_{\text{bot}} = 148$ nm (the second set). We simulate the 6 periods-long systems that correspond approximately to a 2-μm-long waveguide. The reflection spectra are shown in Fig. 10. By applying voltage both reflection maxima grow up in 2-3 times and shift towards higher frequencies. For the first set of parameters there is the enhancement of reflection near 177.5 THz, however, the maximum reflection is shifted towards 200 THz. For the second set one additional peak at 210 THz is clearly seen. The reason for such complex response of reflection on voltage is that waves with propagation constants 10.0 μm$^{-1}$ and 8.9 μm$^{-1}$ are excited also on higher frequencies for both branches "on+" and "on–", as shown in Fig. 4. Therefore, the zone of enhanced reflection is extended from 160 THz to 220 THz.

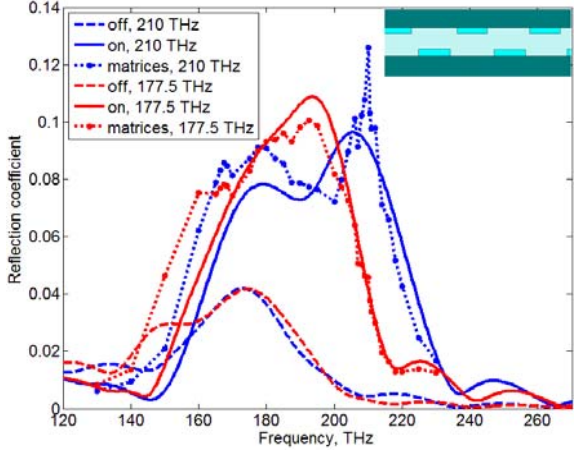

Fig. 10. Reflection coefficients for the Bragg grating systems for enhanced reflection at 177.5 THz ($w_{top}$ = 157 nm and $w_{bot}$ = 177 nm) and at 210 THz ($w_{top}$ = 152 nm and $w_{bot}$ = 148 nm). Thickness of the stripes is 8 nm and total structure thickness between silver plates is 78 nm. Inset shows schematic view of the structure.

To cross-check the phase modulation performance we calculate reflection of the system by the scattering matrix formalism (see e.g. [43]). For the on-state case we assume a sequence of parts with thicknesses $w_{top}$ and $w_{bot}$ and complex propagation constants $\beta_{on+} + i\alpha_{on+}$ and $\beta_{on-} + i\alpha_{on-}$. The effective mode indices are defined as $n_+ = (\beta_{on+} + i\alpha_{on+})/k_0$ and $n_- = (\beta_{on-} + i\alpha_{on-})/k_0$, where $k_0 = 2\pi\nu/c$ is the wave number in vacuum. The whole transmission matrix is a product of transmission matrices for single interfaces and propagation matrices of single layers. The single interface transmission and reflection coefficients in the transmission matrix are defined as $t = 2\sqrt{n_+ n_-}/(n_+ + n_-)$ and $r = (n_+ - n_-)/(n_+ + n_-)$.

Values of $\beta_{on+}$, $\beta_{on-}$, $\alpha_{on+}$ and $\alpha_{on-}$ are taken as the first mode solutions of the SPP dispersion equation for the four-layer system (represented in Fig. 4 and Fig. 5 of Section 4). Reflection spectra calculated with the transfer matrix approach are shown in Fig. 10 with the legend sign "matrices". The agreement with the finite element method simulations is both qualitative and quantitative. Together with the results of Section 5, it validates the approximation of the waveguide as a sequence of parts with different propagation constants. As a result, the 2-μm system designed according to (6) gives higher and broader reflection peaks under applied voltage. Further increase in the number of periods, unfortunately, does not improve the reflection. The reason is that losses in the longer system compensate the enhancing of the Bragg grating reflection.

In principle, such changes in reflection should influence the transmission coefficient. However, actual transmission changes are quite small because of low values of the reflection coefficients. Nevertheless, the Bragg grating configuration with modulation of reflection bears some potential for active elements.

## 7. Discussions and conclusions

Current state-of-the-art plasmonic modulators outperform among Si-based electro-optic modulators [6,21]. However, plasmonic modulators suffer very high propagation losses. Losses are lower in MOS waveguide-integrated designs at the expenses of larger dimensions. The structure which contains an active layer sandwiched between two metal plates allows more compact layout.

We have studied optimization options for the plasmonic absorption modulator reported in [22], which consists of a metal-insulator-metal waveguide with an ITO controlling layer. The reference system with the continuous ITO film has 2.2 dB/μm extinction ratio and 7.2 dB/μm propagation losses. The main target is to increase transmittance. The proposed improvements are based on changes in the ITO permittivity, which can be realized under different fabrication conditions, and periodic patterning of the ITO layer. First, varying of the ITO layer permittivity (with the fixed doping level and change of the carrier density) increases the modulation depth of the device; specifically the FoM rises from 0.30 to 0.37. In particular, optimized system has 3.2 dB/μm extinction ratio and 8.7 dB/μm propagation losses. Thus, the 3-dB modulation can be achieved with the 1-μm-long and ~0.5-μm-wide device. Second, the patterned device exhibits an improved functionality. We have elaborated a theoretical model to explain the transmission and FoM features. Losses of the modulator with periodic ITO stripes are significantly decreased in comparison with the reference modulator design with the continuous ITO layer. To illustrate this improvement we have calculated the transmission coefficients and shown that they increase from 12% (for the reference case) up to 70% in the voltage-off state of the structure with $L = 2$ μm without any significant drop in the FoM. It means that less than 0.8 dB/μm propagation losses can be achieved by patterning of the ITO layer, keeping extinction ratio at the 3 dB/μm level. Thus, our design is competing with the Ag-SiO$_2$-Si-Ag modulator with thicker core (180 nm, see [29]).

Meanwhile we have found some peculiarities in the transmission spectra in dependence on the width of the stripes. For the narrow stripes we explain the pronounced transmission spectra minima by the excitation of nanorod plasmonic resonances. We observe good correlation between simulations and analytic results. The alternative modulation configuration has been considered through utilization of the phase control of the modulator's mode. By elaborating periodic patterning similar to the Bragg grating we have achieved the 2-3 times grow in the maximum of reflection coefficient under applied voltage.

In perspective we would like to point out that not all the resources for the modulator optimizations have been committed. In this work we utilize results from the Thomas-Fermi screening theory. Accordingly to it the carrier density was changed from $9.25 \cdot 10^{26}$ m$^{-3}$ to $9.7 \cdot 10^{26}$ m$^{-3}$ under the 12 V applied bias [22]. The modulator's performance can be drastically enhanced if we take into account possible concentration changes reported in other references. For example, results of [20] are based on the concentration increase from $4.19 \cdot 10^{26}$ m$^{-3}$ to $7.08 \cdot 10^{26}$ m$^{-3}$ (1 V bias is mentioned). Moreover, in the recent work [21] the authors analyze the dramatic almost 70-fold change in the concentration from $0.1 \cdot 10^{26}$ m$^{-3}$ to $6.8 \cdot 10^{26}$ m$^{-3}$ (up to 4 V in experiments). The final mode index change also strongly depends on the concentration averaging over the ITO layer. We average over the whole 8-nm-thick layer. Half-layer averaging (similar to [20]) or even more fine resolution give higher changes in the permittivity. Another potential direction of optimization is connected with the resonances, e.g. an SPP resonance [22] or Fabri-Perot resonance in finite-length metal-dielectric-metal cavities [1, 51], and provides additional advantages in transmission. As a drawback, resonances restrict the operational bandwidth.

**Acknowledgments**

A.L. acknowledges partial financial support from the Danish Research Council for Technology and Production Sciences via the THz COW project.